\documentclass[]{elsart}  
\usepackage{graphicx}
\usepackage{amssymb}
\usepackage{amsmath}
\usepackage{latexsym}
\def\defeq{\mathop{\stackrel{\mathrm{def}}{=}}\nolimits}  
\def\today{21 October 2004} 

\journal{Annals of Physics}

\begin{document}

\begin{frontmatter}

\title{\vspace*{-25pt}Matched detectors as definers of force\\[-40pt]}  

\author[madjid]{F. Hadi Madjid} and 
\author[myers]{John M. Myers}

\address[madjid]{82 Powers Road, Concord, Massachusetts 01742}   

\address[myers]{Gordon McKay Laboratory, Division of Engineering and
Applied Sciences, Harvard University, Cambridge, Massachusetts 02138}

\begin{abstract}
Although quantum states nicely express interference effects, outcomes
of experimental trials show no states directly; they indicate
properties of probability distributions for outcomes. We prove
categorically that probability distributions leave open a choice of
quantum states and operators and particles, resolvable only by a move
beyond logic, which, inspired or not, can be characterized as a guess.

By recognizing guesswork as inescapable in choosing quantum states and
particles, we free up the use of particles as theoretical inventions
by which to describe experiments with devices, and thereby replace the
postulate of state reductions by a theorem.  By using the freedom to
invent probe particles in modeling light detection, we develop a
quantum model of the balancing of a light-induced force, with
application to models and detecting devices by which to better
distinguish one source of weak light from another.  Finally, we
uncover a symmetry between entangled states and entangled detectors, a
dramatic example of how the judgment about what light state is
generated by a source depends on choosing how to model the detector of
that light.
\end{abstract}

\end{frontmatter}

\section{Introduction}\label{sec:1}

Connecting the bench of experiment and the blackboard of theory offers
physicists opportunities for creativity that we propose to make
explicit.  Traditional views underplay the physicist's role in making
these connections.  Although physicists have wished for mathematics
that would connect directly to experiments on the bench, the equations
of quantum mechanics express quantum states and operators not directly
visible in spectrometers or other devices.  Here we look into quantum
mechanics as mathematical language used to model behaviors of devices
arranged on the laboratory bench.  After separating models as
mathematical objects from any assertion that a certain model describes
a given experiment with devices, we ask: given a certain form of
model, which models, if any, fit the behavior of some particular
devices on the bench?  In contrast to any hope for a seamless, unique
blackboard description of devices on a laboratory bench, we argue,
based on mathematical proofs presented in Sec. 3, that no matter what
experimental trials are made, if a quantum model generates calculated
probabilities that match given experimentally determined relative
frequencies, there are other quantum models that match as well but
that differ in their predictions for experiments not yet performed.

The proofs demonstrate what before could only be suspected:
between the two pillars of calculation and measurement must stand a
third pillar of choice making, involving personal creativity beyond
logic, so there can be no reason to expect or demand that any two
people choose alike. 

What does recognizing choice mean for physicists?  In physics, as in
artistic work, pleasure and joy come from the choices one makes that
lead to something interesting.  Looking back, physicists can hardly
help noticing that their proudest accomplishments, whether theoretical
or experimental, have involved choices made by reaching beyond logic
on the basis of intuition, hunches, analogies---some kind of guess,
perhaps inspired, but still outside of logic.  To understand the proofs
is to see opportunities for making guesses.  Although
hunches and guesses and intuition can be as personal as dreams,
the recognition of guesswork as a permanent pillar of physics has
more than personal impact:
\begin{enumerate}
\item Describing device behavior will be recognized in Sec.~4 as a
bi-lingual enterprise, with a language of wrenches and lenses for
experimental trials on the bench and a different language of states
and operators for the blackboard, linked by metaphors as guesswork.
We show how freedom to choose particles as constituents of models of
devices both helps in modeling devices and allows us to replace a
widespread but questionable postulate of ``state reductions'' by a
theorem.
\item The need for bi-lingual descriptions bridging bench and
blackboard gives local color to certain concepts.  In Sec. 5 we
develop a notion of {\em force} in the context of light detection that
gives meaning both at the blackboard and the bench not only to
expectation values of light forces, but also to higher-order
statistics associated with them, with application to models and
detecting devices by which to better distinguish one source of weak
light from another.
\item
In Sec. 6 we uncover a symmetry, pertaining to entanglement, to make
vivid the way judgments about how to model light are interdependent
with judgments about how to model light-detectors.
\end{enumerate}
\vfil

\section{Models mathematically distinct from experiments}\label{sec:2}

Experimental records can hold: (1) numerals interpreted as the
settings of knobs that control an experiment,  and (2) numerals
interpreted as experimental outcomes, thought of as the clicks and
flashes and electronically tallied pulses by which the devices used in
the laboratory respond to and measure whatever is at issue.  As an
abstraction by which one can model experimental outcomes, quantum
theory offers what we shall call {\em theoretical outcomes} (referred
to in the literature variously as {\em outcomes, results, the finding
of particles} and {\em the finding of states}).  Probabilities
of theoretical outcomes are expressed in terms of states and operators
by what we shall call {\em quantum models}.  We discuss 
these first, and then distinguish the probabilities expressed
by models from relative frequencies of experimental outcomes.

\subsection{Definition of quantum-mechanical models}\label{subsec:2.1}
The propositions about linking models to devices (Sec. 3) can be
proved using any formulation of quantum mechanics that includes
probabilities of theoretical outcomes.  Here is a standard formulation
taken from Dirac \cite{dirac} and von Neumann \cite{vN}, bulwarked by
a little measure theory \cite{rudin}; however, as discussed in Sec.\
\ref{sec:4}, we invoke no postulate of state reductions.

Let $H$ be a Hilbert space, let $\rho$ be any self-adjoint
operator of unit trace on $H$ (otherwise known as a density operator),
and let $M$ be a $\sigma$-algebra of subsets of a set $\Omega$ of
possible \textit{theoretical outcomes}.  By a \textit{theoretical
outcome} we mean a number or a list of numbers as a mathematical
object, in contrast to an experimental outcome extracted from an
experimental record.  Let $E$ be any projective resolution on $M$ of
the identity operator on $H$ (which implies that for any $\omega \in
M$, $E(\omega)$ is a self-adjoint projection \cite{rudin2}). Let
$U(t)$ be a unitary time-evolution operator (typically defined by a
Schr\"odinger equation or one of its relativistic
generalizations). These mathematical objects can be combined to define
a probability distribution $\mu$ on $M$, parameterized by $t$:
\begin{equation}\mu(t)(\omega) \stackrel{\rm def}{=}
\mathrm{Tr}[U(t)\rho U^\dag(t)E(\omega)],
\label{eq:mu} \end{equation} where $\mu(t)(\omega)$ is the probability of
an outcome in the subset $\omega$ of $\Omega$, for the parameter value
$t$.

For a probability $\mu(t)(\omega)$ to be compared to anything
experimental, one needs to make explicit the dependence of the $\rho$
and $E$ that generate it on the experimentally controllable
parameters.  It is convenient to think of these parameters as the
settings of various knobs; to express them we let $A$ and $B$ be
mathematically arbitrary sets (interpreted as sets of knob
settings). Let $\mathcal{D}$ be the set of functions from $A$ to
density operators acting on $H$.  Let $\mathcal{E}$ be the set of
functions from $B$ to projective resolutions of the identity on $M$ of
the identity operator on $H$.  Then what we shall call {\em a specific
quantum model} is a triple of functions $(\rho,E,U)$ together
with $\rho \in \mathcal{D}$, $E \in \mathcal{E}$, and $U(t)$ a unitary
evolution operator.  By the basic rule of quantum mechanics, such a
specific quantum model generates a probability-distribution as a
function of knob settings and time: 
\begin{eqnarray}
&&(\forall\ a \in A,\ b \in B,\ t \in I,\ \omega \in
M)\nonumber\\[3pt]
&&\qquad\mu(a,b,t) (\omega)=\mathrm{Tr}[U(t)
\rho(a)U^\dag(t)E(b)(\omega)].
\label{eq:muab} \end{eqnarray} 

Often one needs something less specific than a triple $(\rho,E,U)$.
By a {\em model} we shall mean a set of properties of that limit but
need not fully specify $\rho$, $U$, and~$E$.  For example, in modeling
entangled light, we might construct a model in this sense by specifying
relevant symmetry properties of $E$, $U$, and $\rho$, leaving many
fine points unspecified.

\noindent\textbf{Remarks}
\begin{enumerate}
\item An element $a \in A$ can be a list: $a=\{a^{(1)},\dots
,a^{(n)}\}$; similarly an element $b \in B$ can be a list.
\item
By the {\em domain} of a model, we mean the cartesian product of
the sets $A \times B \times I \times M$.  Models of a given domain
can differ as to the functions $\rho$, $E$, and $U$ defined on the
sets $A$, $B \times M$, and $I$, respectively, so that different
models with a given domain can differ in the states, operators, and
probabilities that they assert.
\item
In defining the domain of a model, we view $\mu$ as a function from $A
\times B \times I \times M \rightarrow [0,1]$, but in writing the left
side of Eq.\ (\ref{eq:muab}) we take the alternative view of $\mu$ as a
function $A\times B \times I \rightarrow \{\mbox{probability measures
on }M\}$.
\item
Different specific models can be distinguished by labels: a model
$\alpha$ consisting of $(\rho_\alpha,U_\alpha,E_\alpha)$ generates a
probability function $\mu_\alpha$.  Because $\mu_\alpha$ is invariant
under any unitary transformation applied to all three of
$\rho_\alpha$, $E_\alpha$, and $U_\alpha$, an important feature of
the states asserted by a specific model $\alpha$ is the
unitary-invariant overlap between pairs of them.  The measure of
overlap convenient to Sec.\ \ref{sec:3} is
\begin{equation} \mathrm{Overlap}(\rho_\alpha(a_1),\rho_\alpha(a_2))
\defeq \mathrm{Tr}[\rho_\alpha(a_1)^{1/2}\rho_\alpha(a_2)^{1/2}].
\label{eq:overlap}\end{equation}
We express the difference in state preparations asserted by specific
models $\alpha$ and $\beta$ having the same domain for elements $a_1,
a_2 \in A_\alpha \cap A_\beta$ by
$|\mathrm{Overlap}(\rho_\alpha(a_1),\rho_\alpha(a_2)) -
\mathrm{Overlap}(\rho_\beta(a_1),\rho_\beta(a_2))|$.
\item We call a model $\beta$ a {\em restriction} of a model $\alpha$ if
the domain of model $\beta$ is a subset of the domain of model $\alpha$.
\end{enumerate}

\subsection{Domains of models and of experimental records}
\label{subsec:2.2}
So much for quantum models as mathematical objects; how do we compare
probabilities from these models with results of an experiment with
lasers and lenses and other devices?  First one contrives to view the
experiment as consisting of trials, each for certain settings of some
knobs, yielding at each trial one of several possible experimental
outcomes.  By tallying the experimental outcomes for various knob
settings, one extracts from the experimental record the relative
frequencies of experimental outcomes as a function on a domain of
experimental knob settings and outcomes.

To compare experimental relative frequencies with probabilities
calculated from a model, both viewed as functions on domains of knob
settings and outcome bins, it is necessary to identify the experimental
domain as a subset of the model domain.  This entails associating to
each experimental outcome $c$ some model outcome $\omega_c \in M$.
For the experimental relative-frequency of outcome $c$ for each
setting of knobs $(a,b,t)$ in the experimentally covered subset of $A
\times B \times I$ we write $\nu_r(a,b,t)(c)$; this is the ratio of
the number of trials with knob settings and time $(a,b,t)$ and an
experimental outcome in $c$ to the number of trials with knob settings
and time $(a,b,t)$ regardless of the outcome.  Letting $C$ denote
the set of experimental outcomes, one has
\begin{equation}(\forall\ t \in I,\ a \in A,\ b \in B)\quad
\sum_{c\in C} 
\nu_r(a,b,t)(c) = 1.\label{eq:nudef}
\end{equation}
By virtue of the mapping $c \mapsto \omega_c \subset \Omega$, one can
compare the experimental relative-frequency function $\nu_r$ with the
probability function $\mu_{\alpha}$ asserted by any model~$\alpha$
having a domain containing a subset identified with the domain of the
experimental relative frequencies.  Because of this need for
identification, a choice of model domain constrains the design, or at
least the interpretation, of experimental records to which models of
that domain can be compared. In compensation, committing oneself to
thinking about an experimental endeavor in terms of a particular model
domain makes it possible to:
\begin{enumerate}
\item organize experiments to generate data that can be compared with
models having that domain;
\item express the results of an experiment mathematically without
having to assert that the results fit any particular model;
\item pose the question of whether the experimental data fit one model
having that domain better than they fit another model \cite{ams}.
\end{enumerate}

\section{Choosing a model to fit given probabilities}\label{sec:3}
What can relative frequencies of outcomes of experimental trials of
devices tell us about how to model those devices?  At best, from
relative frequencies of experimental outcomes as a function of knob
settings, as in Eq.\ (\ref{eq:nudef}), one abstracts an approximation
to a probability-distribution function of knob settings.  By ignoring
statistical and other error, we picture an ideal case of arriving at
some $\mu_\alpha(a,b)(\omega)$, but without any further information
concerning the states or operators that generate it.  This raises a
question inverse to the text-book task of calculating probabilities
from states and operators: given the probability function
$\mu_\alpha(a,b)(\omega)$, what states and operators generate it via
the rule of Eq.\ (\ref{eq:muab})?  Put another way, what are the
constraints and freedoms on a density-operator function $\rho$, an
evolution operator $U(t)$, and a measurement-function $E$ if these are
to generate a given probability function $\mu_\alpha(a,b)(\omega)$ via
the rule of Eq.\ (\ref{eq:muab})?  Here are some answers.

\subsection{Constraint on density operators}\label{subsec:3.1}

If for some values $a_1$, $a_2$, $b$, and $\omega$ one has
$\mu(a_1,b,t)(\omega)$ large and $\mu(a_2,b,t)(\omega)$ small, then
Eq.\ (\ref{eq:muab}) implies that $\rho(a_1)$ is significantly
different from $\rho(a_2)$. This can be quantified in terms of the
\textit{overlap} of two density operators $\rho(a_1)$ and $\rho(a_2)$
defined in Eq.\ (\ref{eq:overlap}).  

\vspace{5pt}

\noindent\textbf{Proposition 1}:\ \ For a specific model $(\rho,E,U)$ to
be consistent with a probability function $\mu$ in the sense of Eq.\
(\ref{eq:muab}), the overlap of density operators for distinct knob
settings $a_1$ and $a_2$ has an upper bound given by
\begin{eqnarray}
\lefteqn{(\forall\ a_1, a_2 \in A)}\quad\nonumber\\[3pt]
&&\mathrm{Tr}[\rho(a_1)^{1/2}\rho(a_2)^{1/2}]\leq \min_{b,t,\omega}
\{[\mu(a_2,b,t)(\omega)]^{1/2} +
 [1-\mu(a_1,b,t)(\omega)]^{1/2}\}.\nonumber\\ 
\label{eq:upper}
\end{eqnarray}

\noindent\textit{Proof}:\ \ For purposes of the proof, abbreviate
$\rho(a_1)$ by $a_1$, $\rho(a_2)$ by $a_2$, and
$U^\dag(t)E(b)U(t)(\omega)$ by~$E$. Because $E$ is a projection, $E =
E^2$.  Then the Schwarz inequality\footnote{For any operators $F$ and
$G$ for which the traces exist, $|\mathrm{Tr}(FG^\dag)| \leq
[\mathrm{Tr}(FF^\dag)]^{1/2} [\mathrm{Tr}(GG^\dag)]^{1/2}$.} and a
little algebra implies
\begin{eqnarray} |\mathrm{Tr}(a_1^{1/2}a_2^{1/2})| &=&
|\mathrm{Tr}(a_1^{1/2}Ea_2^{1/2}) +
 \mathrm{Tr}(a_1^{1/2}(1-E)a_2^{1/2})| \nonumber \\[3pt] & \leq &
 |\mathrm{Tr}(a_1^{1/2}Ea_2^{1/2})| +
 |\mathrm{Tr}(a_1^{1/2}(1-E)a_2^{1/2})| \nonumber \\[3pt] 
\noalign{\goodbreak}
& \leq &
 (\mathrm{Tr}\,a_1)^{1/2}
[\mathrm{Tr}(Ea_2^{1/2}a_2^{1/2}E)]^{1/2}\nonumber\\[3pt]
&&\mbox{} +
 [\mathrm{Tr}(a_1^{1/2}(1-E)(1-E)a_1^{1/2})]^{1/2}
 (\mathrm{Tr}\,a_2)^{1/2} \nonumber \\[3pt] 
&= &
 [\mathrm{Tr}(Ea_2^{1/2}a_2^{1/2}E)]^{1/2} +
 [\mathrm{Tr}(a_1^{1/2}(1-E)(1-E)a_1^{1/2})]^{1/2} \nonumber \\[3pt] &=&
 [\mathrm{Tr}(a_2E)]^{1/2} + [ 1 - \mathrm{Tr}(a_1E)]^{1/2}
 . \end{eqnarray} 
Expanding the notation, we have 
\begin{eqnarray}
&&(\forall\ a_1, a_2 \in A)\nonumber\\[3pt]
&&\qquad\mathrm{Tr}[\rho(a_1)^{1/2}\rho(a_2)^{1/2}]\leq
 \min_{b,t,\omega}\{[\mathrm{Tr}(\rho(a_2)U^\dag(t)E(b)(\omega)U(t))]^{1/2}
 \nonumber \\[3pt] &&\hskip1.85in\mbox{} + 
 [1-\mathrm{Tr}(\rho(a_1)U^\dag(t)E(b)(\omega)U(t))]^{1/2}\},
\end{eqnarray}
 which, with Eq.\ (\ref{eq:muab}), completes the proof. $\Box$

\noindent\textit{Example}:\ \ For $0 \leq \epsilon, \delta \ll 1$, if for
some $b$ and $\omega$, $\mu(a_2,b,t)(\omega) = \epsilon$ and
$\mu(a_1,b,t)(\omega)$ $= 1 - \delta$, then it follows that
$\mathrm{Tr}[\rho(a_1)^{1/2}\rho(a_2)^{1/2}] \leq \epsilon^{1/2} +
\delta^{1/2}$.  If, in addition, $\rho(a_1) = |a_1\rangle\langle a_1|$
and $\rho(a_2) = |a_2\rangle\langle a_2|$, then we have $|\langle
a_1|a_2\rangle|^2 \leq \epsilon^{1/2} + \delta^{1/2}$.

\subsection{Freedom of choice for density operators}\label{subsec:3.2}

The preceding proof of an upper bound invites the question: is there a
corresponding positive lower bound?  The answer turns out to be ``no.''

\vspace{5pt}
\noindent\textbf{Proposition 2}:\ \ For any specific model
$(\rho_\alpha,E_\alpha,U_\alpha)$ and any knob settings $a_1, a_2 \in
A$, regardless of Overlap$(\rho_\alpha(a_1),\rho_\alpha(a_2))$ there
is a specific model $(\rho_\beta,E_\beta,U_\beta)$ with $\mu_\beta =
\mu_\alpha$ and Overlap$(\rho_\beta(a_1),\rho_\beta(a_2)) = 0$.
\vspace{5pt}

\noindent\textit{Proof by construction}:\ \
Let $H_\beta$ be the direct sum of three Hilbert spaces
$H_0$, $H_1$, and $H_2$, each a copy of the Hilbert space $H_\alpha$ of
model $\alpha$: $H_\beta = H_0 \oplus H_1 \oplus H_2$.  Let
$E_\beta(b)(c)$ be the direct sum of three copies of $E_\alpha(b)(c)$,
one for each of the $H_j$; similarly, let $U_\beta$ be the direct sum
of three copies of $U_\alpha$. Define $\rho_\beta$ by
\begin{equation} \rho_\beta(a) = \begin{cases} 
\ \rho_\alpha \oplus 0 \oplus 0 &\mbox{\ if \ }a \neq a_1 \mbox{\ and\ }
a \neq a_2; \\
\ 0 \oplus \rho(a_1) \oplus 0  &\mbox{\ if \ }a = a_1;\\
\ 0 \oplus 0 \oplus \rho(a_2) &\mbox{\ if \ }a = a_2.\end{cases}
\end{equation}
This defines a model $\beta$ of the form of Eq.\ (\ref{eq:muab})
for which we have
\begin{equation}(\forall\ a \in A,\ b \in B,\ t \in I,\ \omega \in M) 
\hskip1em \mu_\alpha(a,b,t) (\omega) = \mu_\beta(a,b,t) (\omega),
\end{equation}
but for the Overlap as defined in Eq.\ (\ref{eq:overlap}),
$\mathrm{Overlap}(\rho_\beta(a_1),\rho_\beta(a_2)) = 0$,
regardless of the value of
$\mathrm{Overlap}(\rho_\alpha(a_1),\rho_\alpha(a_2))$. $\Box$

This proof shows the impossibility of establishing by experiment a
positive lower bound on state overlap without reaching outside of
logic to make an assumption, or, to put it baldly, to {\em guess}
\cite{ams}.  The need for a guess, no matter how educated, has the
following interesting implication. Any experimental demonstration of
quantum superposition depends on showing that two different settings
of the $A$-knob produce states that have a positive overlap.  For
example, a superposition $|a_3\rangle = (|a_1\rangle +
e^{i\phi}|a_2\rangle)/\sqrt{2}$ has a positive overlap with state
$|a_1\rangle$.  Because, by Proposition 2, no positive overlap is
experimentally demonstrable without guesswork, we have the following:

\vspace{5pt}
\noindent\textbf{Corollary to Proposition 2}:\ \ Experimental
demonstration of the super\-position of states requires resort to
guesswork.

\subsection{Constraint on resolutions of the identity}\label{subsec:3.3}

Much the same story of constraint and freedom holds for resolutions of
the identity.  For the norm of an operator $A$ we take $\|A\| =
\sup_u\|Au\|$, where $u$ ranges over all unit vectors. Then we have:

\noindent\textbf{Proposition 3}:\ \ In order for a specific model
$(\rho,E,U)$ to generate a given probability function
$\mu$, the resolution of the identity must satisfy the constraint
\begin{displaymath}
\| E(b_1)(\omega_c) -
E(b_2)(\omega_c) \| \ge \max_{a \in
A,\,t \in I}|\mu(a,b_1,t)(c) - \mu(a,b_2,t)(c)|.
\end{displaymath}

\noindent\textit{Proof}:\ \ Replacing unit vectors by density operators
in the definition of the norm results in the same norm, from which we
have 
\begin{eqnarray*}
\| E(b_1)(\omega_c) - E(b_2)(\omega_c) \|&\ge&
\max_{a \in A,\,t\in
I}|\mathrm{Tr}[U(t)\rho(a)U^\dag(t)
E(b_1) (\omega_c)]\\[3pt] &&\mbox{} -
\mathrm{Tr}[\rho(a)E(b_2)(\omega_c)]|\\[5pt] &
=&\max_{a \in A,\,t \in I}|\mu(a,b_1)(c) - \mu(a,b_2)(c)|.\\
&&\hskip2.9in \Box
\end{eqnarray*}

\subsection{Freedom of choice for resolutions of the identity}
\label{subsec:3.4}
The difference of any two commuting projections has norm less than or
equal to 1.  Can requiring a model $\alpha$ to generate a given
probability function $\mu$ impose any upper bound less than 1 on
 $\|E_\alpha(b_1)(\omega_c) - E_\alpha(b_2)(\omega_c)\|$? 

\noindent\textbf{Proposition 4}: For any specific model
$(\rho_\alpha,E_\alpha,U_\alpha)$ and any knob settings $b_1$ and
$b_2$, regardless of $\|E_\alpha(b_1)(\omega_c) -
E_\alpha(b_2)(\omega_c)\|$ there is a specific model
$(\rho_\beta,E_\beta,U_\beta)$ with $\mu_\beta = \mu_\alpha$ and
$\|E_\beta(b_1)(\omega_c) - E_\beta(b_2)(\omega_c)\| = 1$.

\vspace{5pt}

\noindent\textit{Proof by construction}:\ \ Let the Hilbert space for
model $\beta$ be $H_\beta = H_\alpha \oplus H_\perp$ where $H_\perp$
is a space orthogonal to $H_\alpha$.  Let $E_\beta(b_1)(\omega_c) =
E_\alpha(b_1)(\omega_c)\oplus \mathbf{1}_\perp$ and
$E_\beta(b_2)(\omega_c) = E_\alpha(b_2)(\omega_c)\oplus
\mathbf{0}_\perp$, where $\mathbf{1}_\perp$ is the unit operator on
$H_\perp$ and $\mathbf{0}_\perp$ is the zero operator on $H_\perp$.
Let $\rho_\beta = \rho_\alpha \oplus \mathbf{0}_\perp$.  Then,
$\|E_{\alpha}(b_1)(\omega_c)- E_{\alpha}(b_2)(\omega_c)\|$\break $ = 1$.
$\Box$

\vspace{5pt}

\noindent\textbf{Remark}:
Suppose, as displayed in the proof of Proposition 2, models $\alpha $
and $\beta$ have the same domain and $\mu_\alpha$ = $\mu_\beta$, but
the $\alpha$-states have overlaps differing from those of the
$\beta$-states.  Then there is always a resolution of the identity $E'$
outside the range of $E_\alpha$ and $E_\beta$, such that for some $a$,
$\,\mbox{Tr}[\rho_\alpha(a)E'] \ne \mbox{Tr}[\rho_\alpha(a)E']$.  In
this sense the models $\alpha$ and $\beta$ conflict concerning their
predictions \cite{JOptB}.

\section{Impact on quantum physics}\label{sec:4}
The connection of any specific quantum model to experiments is via a
probability function.  This and the proofs of Propositions 2 and 4
show something that experiments cannot show, namely that modeling an
experiment takes guesswork, and that a model, once guessed, is subject
to surprises arising in experiments not yet performed.  Some guesses
get tested (one speaks of {\em hypotheses}), but testing a guess
requires other guesses not tested.  By way of example, to guide
the choice of a density operator by which to model the light emitted by
a laser, one sets up the laser, filters, and a detector on a bench to
produce experimental outcomes.  But to arrive at any but the
coarsest properties of a density operator one needs, in addition to these
outcomes, a model of the detector, and concerning this model, there must
always be room for doubt; we can try to characterize the detector better,
but for that we have to assume a model for one or more sources of light.
When we link bench and blackboard, we work in the high branches of a tree
of assumptions, holding on by metaphors, where we can let go of one
assumption only by taking hold of others.  Because of the guesswork
needed to bridge between models and experiments, describing device
behavior is forever a bi-lingual enterprise, with a language of wrenches
and lenses for the bench and a different language of states and operators
for the blackboard.

We will show how some words work as metaphors, straddling bench and
blackboard, where by `words' we mean to include whatever mathematical
symbols are used to describe devices.  We consider the mathematics of
quantum mechanics not in contrast to words but as blackboard language,
words of which are sometimes borrowed for use at the bench to describe
devices.  By showing some choices of metaphorical uses of the words
{\em state, operator, spacetime, outcome, and particle}, we promote
freedom to invent particles as needed to describe interesting
features of device behavior.  Recognizing choices in word use reflects
back on how we formulate quantum mechanics: the notion of repeated
measurements `of a state' will be revealed as neither necessary nor
sensible, and the so-called postulate of state reductions will evaporate,
leaving in its place a theorem.

\subsection{Word use at blackboard and lab bench}\label{sec:4.1} 

We start by looking at several related but distinct uses of {\em
spacetime coordinates}.  In the laboratory one uses clocks and rulers
to assign coordinates to acts of setting knobs, transmitting signals,
recording detections, {\em etc}, and one thinks of these
experimentally generated coordinates as points of a
spacetime---something mathematical.  We call this a `linear spacetime'
to distinguish it from a second spacetime, that we call `cyclic,' onto
which the linear spacetime is folded, like a thread wound around a
circle, so that experimental outcomes for different trials can be
tallied in bins labeled by coordinates.  Distinct from linear and
cyclic spacetimes, any quantum-mechanical model involves a third
spacetime on which are defined solutions of a Schr\"odinger equation
(or one of its relativistic generalizations), and it is with
reference to this spacetime that particles as theoretical constructs
are defined.

Any quantum model written in terms of particles generates probabilities,
and if the probabilities of the model fit the relative frequencies of
experimental outcomes well enough, one is tempted to say that one has
``seen the particles''; however, because particles in their mathematical
sense are creatures of models, and multiple, conflicting models are
consistent with any given experimental data, this ``seeing of particles''
stands on guesswork and metaphor, needed, for example, to bind the {\em
electron} as a solution of the Dirac equation defined on a model
spacetime to a flash from a phosphor on a screen. This metaphorical role
of {\em electron, photon, etc.}, though habitual and easily overlooked,
can be noticed when a surprise prompts one to make a change in the use of
the word {\em electron} at the blackboard while leaving the use at the
bench untouched, or {\em vice versa}.\looseness=-1

Next we address notions of (a) components of a theoretical outcome, (b) a
distinction between signal particles and probe particles, and (c)
various measurement times.  We take these in order.

\subsubsection{Multiple components of an outcome}
The term {\em theoretical outcome} pertains to a vector space of
multi-particle wave functions defined on a model spacetime.  This vector
space is a tensor product of factors, one factor for each particle. For a
resolution of the identity that factors accordingly, we shall view each
theoretical outcome for this resolution as consisting of a list of
components, one component for each of the factors.  A probability
density for such multi-component outcomes can be viewed as a joint
probability density for the component parts of the outcome, modeling the
joint statistics of the detection of many particles.

\subsubsection{Signal and probe particles}
We who model are always free to shift the boundary between states (as
modeled by density operators) and measuring devices (as modeled by
resolutions of the identity) so as to include more of the measuring
devices within the scope of the density-operator part of the model
\cite{vN}. Consider for example a coarse model $\alpha$ that portrays a
detecting device by a resolution of the identity.  While a resolution of
the identity has no innards, detecting devices do.  To model, say, a
photo-diode and its accompanying circuitry, we can replace model
$\alpha$ by a more detailed model $\beta$: the quantum state asserted
by model $\alpha$ becomes what we call a {\em signal} state, a factor
in a tensor product (or more generally a sum of tensor products)
accompanied by factors for one or more {\em probe-particle states}.
According to this model $\beta$, the signal state is measured only
indirectly, via a probe state with which it has interacted, followed
by a measurement of the probe states, as modeled by ``a resolution of the
identity'' that works on the probe factor, not the signal factor.

\subsubsection{Measurement times}

Recognizing probe-states as free choices in modeling clarifies a
variety of times relevant to quantum measurements. For any quantum
model $\alpha$, the form of Eq.\ (\ref{eq:muab}) links an outcome
(whether single- or multi-component) to some point time $t_\alpha$;
however, the use of such a model is to describe an actual or
anticipated experiment, and for this, as described above, one is
always free to choose a more detailed model $\beta$, in which the
state of model $\alpha$ appears as the {\em signal} state that
interacts with a probe state, followed by a measurement of the probe
state at some time $t_\beta$ after the interaction.  Thus model
$\beta$ replaces the point time $t_\alpha$ by a time stretch during
which the signal and probe states interact, thus separating the time
during which the signal state interacts with the probe from the time
at which the probe state is measured.

In more complex models involving more probe states, a
succession of ``times of measurement'' in the sense of interactions
can be expressed by a single resolution of the identity.
Finally, in modeling spatially dispersed signal states that interact with
entangled probe particles \cite{aharonov}, one can notice a prior
``probe-interaction time'' during which the probe particles must
interact with one another, in order to have become entangled.

\subsection{State reduction as a theorem, not a postulate}
\label{subsec:4.2}
Recognizing choice in modeling allows one to sidestep a long-troubling
issue in formulating quantum language.  In logical conflict with
the Schr\"{o}dinger equation as the means of describing time evolution
\cite{tai}, Dirac and other authors introduce {\em state reductions}
by a special postulate that asserts an effect on a quantum state of a
resolution of the identity; allegedly needed to express repeated
measurements of a system.  Once we recognize the modeling freedom to
make signal-probe interactions explicit, we can always replace any
story about devices involving a ``state to be measured repeatedly'' by
a model in which a {\em signal} state interacts with a succession of
probe states, followed by a simultaneous measurement of all the probe
states, as expressed by a single resolution of the identity
and a composite state that incorporates both signal and probe states.
Thus any apparent need for a postulate to do with ``repeated
measurements'' evaporates, and with it the unfortunate appearance
of state reductions in a postulate.

Although inconsistent as a postulate, state reduction still works in
many cases as a trick of calculation, as justified by the following
theorem.

\vspace{5pt}

\noindent\textbf{Theorem}:
Assume any specific model of the form Eq.\ (\ref{eq:muab}), and assume
$\omega = (j,k)$ is an outcome with components $j$ and~$k$. If
$E(b)(j,k)$ is a tensor product $E(b)(j,k) = E_A(b)(j) \otimes E_B(b)$,
then for any density-operator function $\rho(a)$, the joint probability
distribution $\mu(a,b)(j,k)$ induces a conditional probability
distribution for $k$ given $j$ that matches the quantum probability of
$k$ obtained using a ``reduced density operator'' obtained by the
usual rule for state reduction applied to~$\rho$.

\vspace{5pt}

\noindent\textit{Proof}: Streamline notation by suppressing the
dependence on $a$ and $b$, and incorporate $U$ into $\rho$, so that the
relevant form is that of Eq.\ (\ref{eq:mu}).  For any state $\rho$ it
follows from Eq.\ (\ref{eq:mu}) that $\Pr(j,k) = \mbox{Tr}[\rho
(E_A(j)\otimes E_B(k))]$.  The conditional probability of $k$ given
$j$ is defined by Bayes rule \cite{feller}:
\begin{eqnarray}\Pr(k|j) & = & \frac{\Pr(j,k)}{\Pr(j)} = 
\frac{\Pr(j,k)}{\sum_{k'}\Pr(j,k')}
=\frac{\mbox{Tr}[\rho (E_A(j)\otimes E_B(k))]}{\mbox{Tr}[\rho
(E_A(j)\otimes \sum_{k'}E_B(k'))]}.
\end{eqnarray}
By the definition of a resolution of the identity, we have
$\sum_{k'}E_B(k') = 1$; recalling that $E_B(k)$ is a
projection that commutes with $E_A(j)$, one then has
\begin{equation}
\Pr(k|j) = \mbox{Tr}[\rho_{\rm red}E_B(k)],
\end{equation}
for an operator 
\begin{equation}\rho_{\rm red} \stackrel{\rm def}{=} \frac{E_A(j)\rho 
E_A(j)}{\mbox{Tr}[E_A(j)\rho E_A(j)]}.
\end{equation}
This $\rho_{\rm red}$ matches the `reduced density operator' obtained
by the usual rule of state reduction.  Q.E.D.

\vspace{5pt}
\noindent\textbf{Remarks}:  
\begin{enumerate}
\item In case $\rho = |\psi\rangle\langle\psi|$ is a pure state,
then $\rho_{\rm red} = |\psi_{\rm red}\rangle\langle\psi_{\rm red}|$,
where the reduced state 
\begin{equation}|\psi_{\rm red}\rangle = \frac{E_A(j)|\psi
    \rangle}{\|E_A(j)|\psi \rangle\|},
\end{equation}
which is one form of the usual rule for state reduction, but here
obtained by calculation with no need for any postulate.
\item Either of the outcome components can be a composite, so the
theorem applies to cases involving more than two outcome components.
\item In relativistic formulations of quantum mechanics, detections at
spatially separated locations $A$ and $B$ can be modeled by
projections of the form assumed.
\end{enumerate}

\section{Balancing forces in the detection of weak light}\label{sec:5}

Of interest in particle physics, astrophysics, and emerging practical
applications, sources of weak light are characterized experimentally
by the experimental outcomes of detectors \cite{sobo}.  Because
detector outcomes are statistical, trial-to-trial differences in
outcomes can arise both from trial-to-trial irregularity in
the sources and from quantum indeterminacy in their detection.
As we shall see, detecting devices work in two parts,
one of which balances a light-induced force against some reference. By
taking advantage of the freedom to invent probe particles when we
model particle detectors, we are led to a quantum mechanical expression
of force in the context of balancing devices, with application to
models and detecting devices by which to better distinguish one source
of weak light from another.

In Newtonian physics, the word {\em force} is used both on the
blackboard and with balancing devices. In quantum physics, {\em
force}, as used at the blackboard, gets re-defined in terms of the
expectation values pertaining to dynamics of wave functions
\cite{feynmanIII}.  We will find useful a concept of force in
characterizing light.  Because of its employment in experimental
work, our concept of {\em force} necessarily takes on local coloring
special to one or another experimental bench; we develop a notion of
{\em force} in the context of light detection that gives meaning both
at the blackboard and the bench not only to expectation values of
forces, but also to higher-order statistics associated with them.
These higher-order statistics allow the expression, within quantum
mechanics, of the teetering of a balance that happens when forces are
nearly equal. We begin by reviewing some details of detector behavior.

\subsection{Balancing in detectors}\label{subsec:5.1}

Under circumstances to be explored, particle detectors employed to
decide among possible quantum states produce unambiguous experimental
outcomes.  Seen up close, a detecting device consists of two components. 
The first is a {\em transducer} such as a photo-diode that responds to
light by generating a small current pulse.  To tally a transducer
response as corresponding to one or another theoretical outcome in the
sense of quantum mechanics, one has to {\em classify} the response using
some chosen criterion.  As phrased in the engineering language of inputs
and outputs, the response of the transducer is fed as an input to a
second component of the detector, in effect an unstable balance
implemented as a flip flop (made of transistors organized into a
cross-coupled differential amplifier). The flip-flop produces an out{\em
put} intended to announce a decision between two possible experimental
out{\em comes}, say 0 and~1.

If we think classically, we picture the flip-flop as a ball and two
bins, one bin for each possible outcome, separated by a barrier, the
height of which can be adjusted, as shown in Fig.\ \ref{fig:1}.  The
ball, starting in bin 0 is kicked by the transducer; an outcome of 1 is
recorded if and only if the balance is tipped and the ball rolls past the
barrier into bin~1. This ball-and-bin technique avoids ambiguity by
virtue of a convention that gives the record a certain leeway: it does
not matter if the ball is a little off center in its bin, so long as the
ball does not teeter on the barrier between bins.

Although usually producing an unambiguous outcome, the flip-flop, seen
up close, can teeter in its balancing, perhaps for a long time, before
slipping into one bin or the other \cite{gray,chaney}. Absent some special
intervention, two parties (people or machines) to which a teetering
output fans out can differ in how they classify this output as an
outcome: one finds a 0, the other finds a 1.  To reduce the risk of
disagreement, the two parties have to delay their reading of the
output, hopefully until the ball slips into one or the other bin.

Ugly in this classical cartoon are two related features: (1) the ball
can teeter forever, so that waiting is no help, and (2) the mean time
for teetering to end is entirely dependent on some {\em ad hoc}
assumption about ``noise'' \cite{gray,chaney}.  Although we have described
the flip-flop classically, it is built of silicon and glass presumably
amenable to quantum modeling.  To see what quantum models offer us,
the first step is to recall that a quantum model implies probabilities
to be related to an experiment, so that inventing a quantum model and
choosing an experimental design go hand in hand.  Thirty years ago,
thinking not in quantum but in circuit terms, we designed and carried
out an experiment to measure teetering of the output of a 
flip-flop.  As we recognized only recently, the record of this
experiment is compatible with some quantum models.  A quantum
model to be offered shortly describes the experiment already
performed and serves as a guide for designing future experiments to
exploit what can be called a statistical texture of force, previously
obscured by the ``noise'' invoked in classical analysis. 

\subsection{Experimental design}\label{subsec:5.2}

To experiment with the teetering of a detecting device comprised of a
transducer $D_0$ (for diode) connected to a flip-flop
$F_0$, shown in Fig.~\ref{fig:2}, we replace the
transducer $D_0$ by a laboratory generator of weak electrical pulses
to drive $F_0$.\footnote{Rather than a separate transducing
photo-diode as input to $F_0$, one could replace one transistor of
flip-flop $F_0$ by a photo-transistor.}  Putting the flip-flop $F_0$
into a teetering state takes very sensitive adjustment of the pulse
generator, achieved by feedback from a running average of the outcomes
produced by $F_1$ and $F_2$ to the pulse generator. The output of
$F_0$ is made to fan out, as shown in Fig.~\ref{fig:3}, to a matched
pair of flip-flops, $F_1$ and $F_2$, each of which acts as an
auxiliary detector, not of the incoming light but of the output of
$F_0$.  The flip-flops $F_1$ and $F_2$ are clocked at a time $T$ later
than is $F_0$.  The experimental outcome consists of two binary
components, one from $F_1$ and the other from $F_2$.  If after the
waiting time $T$ the output of $F_0$ is still teetering, the
flip-flops $F_1$ and $F_2$ can differ, one registering a 1 while the
other registers a 0; the disagreement between the two flip-flops
registers the teetering of the output of $F_0$.  The measured relative
frequency of disagreements between  $F_1$ and $F_2$ is shown in
Fig.~\ref{fig:4}.

\subsection{Quantum modeling of balancing in detectors}\label{subsec:5.3}

We want to model the teetering statistics that we will use to discriminate
among various sources of weak signals. Traditional analyses of
solid-state detecting devices and their flip-flops invoke quantum
mechanics only to determine parameters for classical stories involving
voltage and current.  Analyzed that way, teetering in a
photo-diode-based detector that employs a flip-flop made of
transistors arises in two ways: first, there can be teetering in the
entry of electrons and holes into the conduction band of the
photo-diode; second, there can be teetering in the response of $F_0$
to whatever amplified pulse comes from the photo-diode.  Although both
these teeterings involve electrons and holes going into a conduction
band, the statistical spread of outputs for a given state is blamed on
{\em noise} unconnected with the signal, and known analyses of a
flip-flop invoke {\em noise} to evade the embarrassment of a possible
infinite hesitation.

Avoiding the invocation of `noise,' we picture $F_0$ quantum
mechanically as a pair of probe particles.  Light acting via the
transducer applies a force to the two probe particles. This scattering
process transforms an initially prepared state of the light and probes
to an out-state consisting of a sum (or integral) of products, each of
which has a factor for the light and a factor for the two probe
particles of $F_0$.  After some waiting time of evolution $T$, the two
probe particles are measured, as expressed by a resolution of the
identity that ignores the signal state; thus the probabilities of
theoretical outcomes of the detection after the interaction are
expressible by a reduced density operator obtained by tracing over the
signal states.  In this view, teetering shows up in the probability of
detecting the two probe particles on different sides of a reference;
we will show how this probability depends on both the signal detected
and a waiting time $T$, and how in this dependence Planck's constant
enters.

In the model presented here, we simplify the effect of the signal
state to that of preparing, at time 0, a pair of probe-particle wave
functions.  For simplicity, the probe-particle wave functions have
only one space dimension. Let $x$ be the space coordinate for one
probe particle and $y$ be the space coordinate for the other.  The
difference between one possible signal state and another is reflected
by concentrating the initial probe-state wave functions
slightly to one side or the other of an energy hump centered at the
origin $x=0$, $y = 0$.  (As might be expected, teetering is most
pronounced in the borderline case of a signal state that puts the
initial probe-particle wave functions evenly over the energy hump.) We
model the flip-flop $F_1$ as a resolution of the identity that has an
theoretical outcome of 1 for $x > 0$ and 0 for $x < 0$; similarly
$F_2$ is modeled as a resolution of the identity for $y$.  Thus the
two-component theoretical outcomes are 00 for $x,y<0$; 01 for $x<0$,
$y>0$; 10 for $x>0$, $y<0$; and 11 for $x,y>0$.  By assuming a
coupling between the two probe particles, we will model how increasing
the waiting until time $T$ to detect the probe particles decreases the
probability of disagreement between $F_1$ and $F_2$, {\em i.e.}\
diminishes the probability of $x$ and $y$ being measured with
different signs.

Thinking of the  $x$-probe particle as a wave-function concentrated
near an energy hump, assuming that the long-time behavior
of the particle depends only on the hump curvature,
and for the moment neglecting coupling between the two probe
particles, we express the dynamics of the $x$-particle by the
Schr\"odinger equation for an unstable oscillator:
\begin{equation}i\hbar\,\frac{\partial \;}{\partial t}\,\psi(x,t) =
\left(-\frac{\hbar^2}{2m}\,\frac{\partial^2 \;}{\partial x^2} -
\frac{k x^2}{2}\right)\psi(x,t),
\label{eq:single}\end{equation} 
where the instability comes from the minus sign in the term
proportional to $x^2$.  We express the $y$ probe particle similarly.  In
order to produce growth over time in the correlation of the detection
probabilities, we put in the coupling term $\frac{1}{4}k\lambda(x-y)^2$. 
This produces the following two-particle Schr\"odinger equation which is
the heart of our model:
\begin{eqnarray}i\hbar\,\frac{\partial \;}{\partial t}\,\psi(x,y,t) &=&
-\frac{\hbar^2}{2m} \left(\frac{\partial^2 \;}{\partial x^2}
+\frac{\partial^2 \;}{\partial y^2}\right)\psi(x,y,t)\nonumber\\
&&\mbox{} +
\frac{k}{2}\left(-x^2 - y^2 + \frac{\lambda}{2}
(x-y)^2\right)\psi(x,y,t). \label{eq:pairphys}
\end{eqnarray}
The natural time parameter for this equation is $\omega^{-1}$ defined
here by $\omega \defeq$\break $\sqrt{k/m}$; similarly there is
a natural distance parameter $\sqrt{\hbar/m\omega}$.  

\subsection{Initial conditions}\label{subsec:5.4} 

{}For the initial condition, we will explore a wave packet of the form:
\begin{equation}
\psi(x,y,0) = \frac{1}{\pi^{1/2}b}\exp[-(x-c)^2/2b^2]\exp[-(y-c)^2/2b^2].
\label{eq:initxy}
\end{equation}
For $c=0$, this puts the recording device exactly on edge, while
positive or negative values of $c$ bias the recording device toward 1 or
0, respectively.

\subsection{Solution}\label{subsec:5.5}

As discussed in Appendix~A, the solution to this model is
\begin{eqnarray}
|\psi(x,y,t)|^2 &=& \frac{1}{\pi B_1(t)B_2(t)}\nonumber\\
&&\mbox{}\times
\exp\left\{-\left(\frac{x+y}{\sqrt{2}} - c\sqrt{2}\cosh
  t\right)^2\Bigg/B^2_1(t)- \frac{(x-y)^2}{2B^2_2(t)}\right\},\nonumber\\
\label{eq:jointxyphys}
\end{eqnarray}
with
\begin{eqnarray}
B_1^2(t) &=& b^2 \left[1 + \left(\frac{\hbar^2}{\omega^2 m^2 b^4}
+1\right)\sinh^2 t\right],
\nonumber \\
B_2^2(t) &=& b^2\left[1 + \left(\frac{\hbar^2}{\omega^2 m^2
b^4(\lambda-1)}-1\right)
  \sin^2 \sqrt{\lambda-1}\,t\right].
\label{eq:Bphysdef}
\end{eqnarray}

The probability of two detections disagreeing is the integral of this
density, $|\psi(x,y,t)|^2$, over the second and fourth quadrants of the
$(x,y)$-plane.  For the especially interesting case of $c =0$, this
integral can be evaluated explicitly as shown in Appendix A:
\begin{eqnarray}
\lefteqn{\Pr(F_1\mbox{ and }F_2\mbox{ disagree at }t)}\qquad\nonumber\\
&=&
\frac{2}{\pi}\tan^{-1}\left(
\frac{\displaystyle 1 + \left[\frac{\hbar^2}{\omega^2 m^2
b^4(\lambda-1)}-1\right]\sin^2\sqrt{\lambda-1}\,\omega t}{\displaystyle 1
+\left(\frac{\hbar^2}{\omega^2 m^2 b^4}+1\right)\sinh^2 \omega
t}\right)^{1/2}.
\label{eq:edge1phys}
\end{eqnarray}
This formula works for all real $\lambda$. For $\lambda > 1$, it shows
an oscillation, as illustrated in Fig.~\ref{fig:4}.  For the case $0 <
\lambda <1$, the numerator takes on the same form as the denominator, but
with a slower growth with time and lacking the oscillation, so that the
probability of disagreement still decreases with time, but more
slowly.  Picking values of $b$ and $\lambda$ to fit the experimental
record, we get the theoretical curve of Fig.~\ref{fig:4}, shown in
comparison with the relative frequencies (dashed curve) taken from the
experimental record.  For the curve shown, $\,\lambda = 1.81$ and $b =
0.556$ times the characteristic distance $\sqrt{\hbar/\omega m}$.
According to this model $\alpha$, a design to decrease the half-life
of disagreement calls for making both $k/m$ and $\lambda$ large.
Raising $\lambda$ above 1 has the consequence of the oscillation,
which can be stronger than that shown in Fig.~\ref{fig:4}.  When the
oscillation is pronounced, the probability of disagreement, while
decreasing with the waiting time $t$, is not monotonic, so in some
cases judging sooner has less risk of disagreement than judging later.

\subsection{An alternative to model $\alpha$}\label{sec:5.6}

As Proposition 2 of Sec.~\ref{sec:3} suggests, one can construct
alternatives to the above model $\alpha$ of a flip-flop.  Instead of
initial probe states specified by ``locating blobs,'' expressed in the
choice of the value of $c$ in Eq.\ (\ref{eq:initxy}), a model $\beta$
can employ initial probe states specified by momenta. In this
``shooting of probe particles at an energy hump,'' the initial wave
functions are concentrated in a region $x,y < 0$ and propagate toward
the energy saddle at $x,y = 0$. Writing a 0 is expressed by an
expectation momentum for the initial state less than that for the
initial state that corresponds to writing a 1.  Hints for this
approach are in the paper of Barton \cite{barton}, which contains a
careful discussion of the energy eigenfunctions for the single
inverted oscillator of Eq.\ (\ref{eq:phiprob}), as well as of wave
packets constructed from these eigenfunctions.  Such a model $\beta$
based on an energy distinction emphasizes the role of a flip-flop as a
decision device: it ``decides'' whether a signal is above or below the
energy threshold.

\subsection{The dependence of probability of disagreement on
$\hbar$}\label{subsec:5.7}

{}For finite $b$, the limit of Eq.\ (\ref{eq:edge1phys}) as $\hbar
\rightarrow 0$ is
\begin{equation}
\Pr(F_1\mbox{ and }F_2\mbox{ disagree at }t) =
\frac{2}{\pi}\tan^{-1}\left(\left|
\frac{\cos \sqrt{\lambda-1}\,\omega t}{\cosh \omega t}\right|\right).
\label{eq:edgeh0}
\end{equation}
This classical limit of model $\alpha$ contrasts with the
quantum-mechanical Eq.\ (\ref{eq:edge1phys}) in how the disagreement
probability depends on $\lambda$.  Quantum behavior is also evident in
entanglement exhibited by the quantum-mechanical model.  At $t = 0$ the
wave function is the unentangled product state of Eq.\
(\ref{eq:initxy}).  Although it remains in a product state when viewed in
$(u,v)$-coordinates discussed in Appendix A, as a function of
$(x,y)$-coordinates it becomes more and more entangled with time, as
it must to exhibit correlations in detection probabilities for the
$x$- and $y$-particles.  By virtue of a time-growing entanglement and
the stark contrast between Eq.\ (\ref{eq:edge1phys}) and its
classical limit, the behavior of the 1-bit recording device exhibits
quantum-mechanical effects significantly different from any classical
description.

The alternative model $\beta$ based on energy differences can be
expected to depend on a \textit{sojourn time} with its interesting
dependence on Planck's constant, as discussed by Barton \cite{barton}.
Both models $\alpha$ and $\beta$ thus bring
Planck's constant into the description of decision and recording
devices, not by building up the devices atom by atom, so to speak, but
by tying quantum mechanics directly to the experimentally recorded
relative frequencies of outcomes of uses of the devices.

\subsection{Balancing and the characterization of light force}
\label{subsec:5.8} 

Detection of teetering in a detector of weak light pulses allows
finer distinctions by which to characterize a source of that light.
Without teetering, a first measure of a weak light source is its mean
intensity, expressed operationally as the fraction of 1's detected in
a run of trials in which it illuminates a detector.  Now comes a
refinement that draws on teetering.  If the detector's balancing
flip-flop $F_0$ fans out to auxiliary flip-flops $F_1$ and $F_2$, two
sources $A$ and $B$ that produce the same fraction of 1's can be
tested for a finer-grained distinction as follows.  For each source,
using feedback to stabilize the relation between the source and $F_0$,
so that the running fraction of 1's detected is held nearly steady,
conduct one run of trials for a fixed waiting time $T_1$, a second run
of trials for a fixed waiting time $T_2$, {\em etc.}  Let $\nu_A(T_k)$
denote the fraction of trials of source $A$ for which the outcome
components from $F_1$ and $F_2$ disagree in the run of trials with
waiting time $T_k$; similarly, $\nu_B(T_k)$ denotes this fraction for
source $B$.  These additional data express additional statistical
``texture'' by which to compare source $A$ against source $B$. Even
when they produce the same overall fractions of 1's, they are still
measurably distinguishable if they consistently show strong
differences, for some $T_k$, between $\nu_A(T_k)$ and $\nu_B(T_k)$.
For example, if source $B$ has more classical jitter than source
$A$, so that the $B$ quantum state bounces up and down in expectation
energy from pulse to pulse, then source $B$ is more apt than source
$A$ to push the probe particles both over the hill or neither over the
hill.  In other words, source $A$ will produce more teeterings, and
hence we would find $\nu_A(T_k)$ significantly greater than
$\nu_B(T_k)$.

\section{Entangled signals {\em vs}.\ entangled balances} 

With the the freedom to invoke probe particles as developed in
Secs.\ 4 and 5, we can show a striking additional freedom of choice
in modeling, resolvable only by going beyond the application of logic
to experimental data.  This freedom pertains to entanglement.

Suppose that experimental trials yield outcomes consistent with a
model, according to which a source of entangled weak light 
illuminates a pair of unentangled detectors at two separate locations,
$A$ and $B$.  Models of detectors detailed enough to invoke
probe-particle states, as in Sec. 5, must specify how these
probes are prepared.  As discussed in a different context long ago
\cite{aharonov}, there is the possibility of entangling the probe
particles, amounting to entangling the detectors.  This possibility of
an entangled pair of detectors points to a symmetry relation between
entanglement of signals and entanglement of probes.

To see how this works, consider first modeling a single detector
involving a probe state prepared by choosing some $|p(q)\rangle $,
where $q$ is a parameter such as the expectation momentum for this
state. Consider also a set of possible signal states $|s(q)\rangle$.
Assume that the detector model calls for detecting the probe particle
after its interaction with the signal particle, as expressed
by a measurement operator $E$ acting on the probe alone.  Then the
probability of outcome $j$ resulting from an initial signal $|s(q)\rangle$
interacting with a probe $|p(q')\rangle $ has the form of the square
of a complex amplitude that depends on both $s(q)$ that labels the
signal state and $p(q')$ that labels the probe state:
\begin{equation}
\Pr(j|s(q),p(q')) =\|\mbox{Amp}(j|s(q),p(q'))\|^2,
\label{eq:pj}
\end{equation}
with 
\begin{equation}
\mbox{Amp}(j|s(q),p(q')) = (1_s \otimes E(j))U (|s(q)\rangle
\otimes |p(q') \rangle).
\end{equation}
Here $U$, acting on the combined signal-probe space of wave functions,
expresses the signal-probe interaction.  Assume for the moment that
the probability in Eq.\ (\ref{eq:pj}) is symmetric under interchange
of signal and probe:
\begin{equation}
\Pr(j|s(q),p(q')) = \Pr(j|s(q')p(q)),
\label{eq:symP}
\end{equation}
which implies that for some real-valued phase-function $\phi$,
\begin{equation}
\mbox{Amp}(j|s(q),p(q')) = e^{i\phi(q,q';j)}\mbox{Amp}(j|s(q')p(q)).
\label{eq:symA}
\end{equation}

To see how this symmetry impacts on modeling a pair of detectors,
consider two such detectors, one at location $A$, the other at $B$,
having identical initial probe states $|p_A(q_0)\rangle$ and
$|p_B(q_0)\rangle$, respectively.  For unentangled signals having
initial states $|s_A(q_1)\rangle$ and $|s_B(q_2)\rangle$, the
amplitude for the joint outcome $j_A$ at $A$ and $j_B$ at $B$ is then
\begin{eqnarray}
\lefteqn{\mbox{Amp}(j_A,j_B|s_A(q_1)s_B(q_2);p_A(q_0)p_B(q_0))=}
\hskip1in\nonumber\\[8pt] & & [(1_{sA}\otimes
E_A(j))\otimes(1_{sB}\otimes E_B(k))](U_A\otimes U_B) \nonumber \\[8pt] 
&&
\quad [(|s_A(q_1)\rangle |p_A(q_0)\rangle)\otimes (|s_B(q_2)\rangle
|p_B(q_0)\rangle)]
 \label{eq:joe} 
\end{eqnarray}
(which can be written as a product of $A$- and $B$-factors, expressing the
lack of correlation when there is no entanglement).  Now consider the
same pair of detectors responding to an entangled signal state
\begin{equation}
\mathcal{N}[s_{A1}s_{B2} + e^{i\theta}
s_{A2} s_{B1}],
\label{eq:entstate}
\end{equation} 
where $\mathcal{N}$ is a normalizing constant, dependent on $q$ and
$q'$, and we have condensed the notation by writing $p_{A1}$ for
$|p_A(q_1)\rangle$, {\em etc}.  The combined signal-probe state written
with the tensor products in the order assumed in Eq.\ (\ref{eq:joe})
is then
\begin{equation}
\mathcal{N}[s_{A1}p_{A0}s_{B2}p_{B0} + e^{i\theta}
s_{A2}p_{A0} s_{B1}p_{B0}];
\label{eq:entstate3}
\end{equation} 
thus the joint probability of outcomes for the entangled signal state is
\begin{eqnarray}\lefteqn{
\Pr(j_A,j_B|\mathcal{N}[s_{A1}s_{B2}+e^{i\theta}s_{A2}s_{B1}],p_{A0}
p_{B0}) =}
\hspace{1.0in} \nonumber \\[8pt] & & \|\mathcal{N}[(1_{sA}\otimes
E_A(j))\otimes(1_{sB}\otimes E_B(k))](U_A\otimes U_B)\nonumber \\[8pt]& &
\;\;[s_{A1}p_{A0}s_{B2}p_{B0} + e^{i\theta} s_{A2}p_{A0}
s_{B1}p_{B0}]\|^2.
\label{eq:joe1}
\end{eqnarray}
Assuming the invariance up to phase of Eq.\ (\ref{eq:symA}), the
exchange of signal and probe parameters results only in
phases that leave the probability unaffected, leading to the relation
\begin{eqnarray}
\lefteqn{\Pr(j_A,j_B|\mathcal{N}[s_{A1}s_{B2}+e^{i\theta}s_{A2}s_{B1}],p_{A0}
p_{B0})}\quad\nonumber\\[8pt]
&=&\Pr(j_A,j_B|s_{A0}s_{B0},\mathcal{N}[p_{A1}p_{B2}+
e^{i\theta}p_{A2}p_{B1}]), 
\end{eqnarray}
so that outcome probabilities are the same for an entangled state
measured by untangled detectors as they are for an unentangled
state measured by entangled detectors.

Without assuming Eqs.\ (\ref{eq:symP}), (\ref{eq:symA}), one can still
ask: given a model $\alpha$ that ascribes probabilities of outcomes to
an entangled signal measured via an unentangled probe state
$p_{A0}p_{B0}$, is there an alternative model $\beta$ that ascribes
the same probabilities to an unentangled signal state measured via an
entangled probe state?  We conjecture that the answer is ``yes,'' in
which case no experiment can distinguish between a model of it that
asserts entangled signal states measured via unentangled probes and a
model that asserts unentangled signal states measured via entangled
probes.

\section{Acknowledgments} 

Tai Tsun Wu called our attention to the conflict between the
Schr\"odinger equation and state reductions.  We thank Howard E. Brandt
for discussions of quantum mechanics.  We thank
Dionisios Margetis for analytic help.

This work was supported in part by the Air Force Research Laboratory and
DARPA under Contract F30602-01-C-0170 with BBN Technologies.

\appendix
\newpage
\section*{Appendix A.\ \ Solution to Model $\alpha$ of a 1-bit recording
device}\label{sec:App}

\newcounter{appendix}
\setcounter{appendix}{1}
\def\theappendix{\Alph{appendix}}
\def\theequation{\theappendix.\arabic{equation}}

Starting with Eq.\ (\ref{eq:pairphys}), and writing $t$ as
the time parameter times a dimensionless ``$t$'' and $x$ and $y$ as
the distance parameter times dimensionless ``$x$'' and ``$y$,''
respectively, we obtain
\begin{equation}i\,\frac{\partial \;}{\partial t}\,\psi(x,y,t) =
\frac{1}{2}
\left(-\frac{\partial^2 \;}{\partial x^2} -\frac{\partial^2
\;}{\partial y^2} - x^2 - y^2 + \frac{\lambda}{2}
(x-y)^2\right)\psi(x,y,t).  \label{eq:pair}
\end{equation}
This equation is solved by introducing a non-local coordinate change:
\begin{equation} u = \frac{x +y}{\sqrt{2}}
\quad\mbox{and}\quad v = \frac{x-y}{\sqrt{2}}.
\end{equation}
With this change of variable, Eq.\ (\ref{eq:pair}) becomes
\begin{equation}i\,\frac{\partial \;}{\partial t}\,\psi(u,v,t) =
\frac{1}{2}\left(-\frac{\partial^2 \;}{\partial u^2} -\frac{\partial^2
\;}{\partial v^2} - u^2 + (\lambda-1) v^2\right)\psi(u,v,t),
\label{eq:uvpair}
\end{equation} for which separation of variables is immediate,
so the general solution is a sum of products, each of the form
\begin{equation}
\psi(u,v,t) = \phi(u,t)\chi(v,t).
\label{eq:factor}
\end{equation}
The function $\phi$ satisfies its own Schr\"odinger  
equation,  
\begin{equation}
\left(i\,\frac{\partial \;}{\partial t} + \frac{1}{2}\,\frac{\partial^2
\;}{\partial u^2} + \frac{u^2}{2}\right)\phi(u,t) = 0,  
\label{eq:phieq}
\end{equation}
which is the quantum-mechanical equation for an unstable harmonic
oscillator, while $\chi$ satisfies
\begin{equation}
\left(i\,\frac{\partial \;}{\partial t} + \frac{1}{2}\,
\frac{\partial^2 \;}{\partial v^2} + \frac{1- \lambda}{2}\,
v^2\right)\chi(v,t)  = 0 , \label{eq:chieq}
\end{equation}
which varies in its interpretation according to the value of
$\lambda$, as follows: (a) for $\lambda < 1$, it expresses an unstable
harmonic oscillator; (b) for $\lambda = 1$, it expresses a free particle;
and (c) for $\lambda > 1$, it expresses a stable harmonic oscillator.  The
last case will be of interest when we compare behavior of the model
with an experimental record.

By translating Eq.\ (\ref{eq:initxy})\ into $(u,v)$-coordinates, one
obtains initial conditions
\begin{eqnarray}
\phi(u,0) &=&
\pi^{-1/4}b^{-1/2}\exp\left\{-\frac{(u-\sqrt{2}c)^2}{2b^2}\right\}, \\
\chi(v,0) &=& \pi^{-1/4}b^{-1/2}\exp\left\{-\frac{v^2}{2b^2}
\right\}.
\end{eqnarray}
The solution to Eq.\ (\ref{eq:phieq}) with these
initial conditions is given by Barton \cite{barton}; we deal with
$\phi$ and $\chi$ in order.  From (5.3) of Ref.~\cite{barton}, one has
\begin{equation}
|\phi(u,t)|^2 = \frac{1}{\pi^{1/2}B_1(t)}\,\exp\left\{\frac{-(u -
c\sqrt{2}\cosh
 t)^2}{B_1^2(t)}\right\},
\label{eq:phiprob} 
\end{equation}
where
\begin{equation}
B_1^2(t) = b^2 \left[1 + \left(\frac{1}{b^4} +1\right)\sinh^2 t\right].
\label{eq:Bdef}
\end{equation}
Similarly, integrating the Green's function for the stable oscillator
($\lambda > 1)$ over the initial condition for $\chi$ yields
\begin{equation}
|\chi(v,t)|^2 =
\frac{1}{\pi^{1/2}B_2(t)}\,\exp\left\{\frac{-v^2}{B_2^2(t)}\right\},
\end{equation}
where
\begin{equation}
B_2^2(t) = b^2\left[1 + \left(\frac{1}{b^4(\lambda-1)}-1\right)
  \sin^2\sqrt{\lambda-1}\,t\right].
\label{eq:B2def}
\end{equation}
Multiplying these and changing back to $(x,y)$-coordinates yield the
joint probability density
\begin{eqnarray}
|\psi(x,y,t)|^2 &=& \frac{1}{\pi B_1(t)B_2(t)}\nonumber\\[8pt]
&&\mbox{}\times \exp\left\{-\left(\frac{x+y}{\sqrt{2}} - c\sqrt{2}\cosh
  t\right)^2\Bigg/ B^2_1(t)- \frac{(x-y)^2}{2B^2_2(t)}\right\}.\nonumber\\
\label{eq:jointxy}
\end{eqnarray}
The probability of two detections disagreeing is the integral of this
density over the second and fourth quadrants of the $(x,y)$-plane.
This is most conveniently carried out in $(u,v)$-coordinates.  For the
especially interesting case of $c =0$ (and $\lambda > 1)$, this
integral can be transformed into
\begin{eqnarray}
\lefteqn{\Pr(F_1\mbox{ and }F_2\mbox{ disagree at
}t)}\qquad\nonumber\\[6pt] &=&
\frac{1}{\pi B_1(t) B_2(t)}\int^{\infty}_{-\infty} dv \int^v_{-v}du\,
\exp\left\{\frac{-u^2}{B_1^2(t)} - \frac{v^2}{B_2^2(t)}\right\} \nonumber
\\[6pt] &=&
\frac{4}{\pi}\int_0^{\infty}e^{-V^2}\, dV\int_0^{B_2(t) V/B_1(t)}dU\,
e^{-U^2} \nonumber \\[6pt] \noalign{\goodbreak} &=&
\frac{2}{\pi}\tan^{-1}\left(\frac{B_2(t)}{B_1(t)}\right)
\nonumber \\[6pt]\noalign{\goodbreak}
 &=& \frac{2}{\pi}\tan^{-1}\left(
\frac{\displaystyle 1 +
\left[\frac{1}{b^4(\lambda-1)}-1\right]\sin^2\sqrt{\lambda-1}\,
t}{\displaystyle 1 +\left(\frac{1}{b^4}+1\right)\sinh^2 t}\right)^{1/2}.\qquad
\label{eq:edge1}
\end{eqnarray}
It is easy to check that this formula works not only when $\lambda >
1$ but also for the case $\lambda < 1$.  For $0 < \lambda <1$,
the numerator takes on the same form as the denominator, but with
a slower growth with time, so that the probability of disagreement
still decreases with time exponentially, but more slowly.

Converting Eq.\ (\ref{eq:jointxy}) from dimensionless back to physical
time and distance variables results in Eq.\ (\ref{eq:jointxyphys})
with Eqs.\ (\ref{eq:Bphysdef}), and similarly Eq.\ (\ref{eq:edge1})
leads to Eq.\ (\ref{eq:edge1phys}).

\newpage
\renewcommand\thefigure{\arabic{figure}}
\setcounter{figure}{0}    

\begin{figure}[t]
\begin{center}
   \includegraphics[width=4.5in]{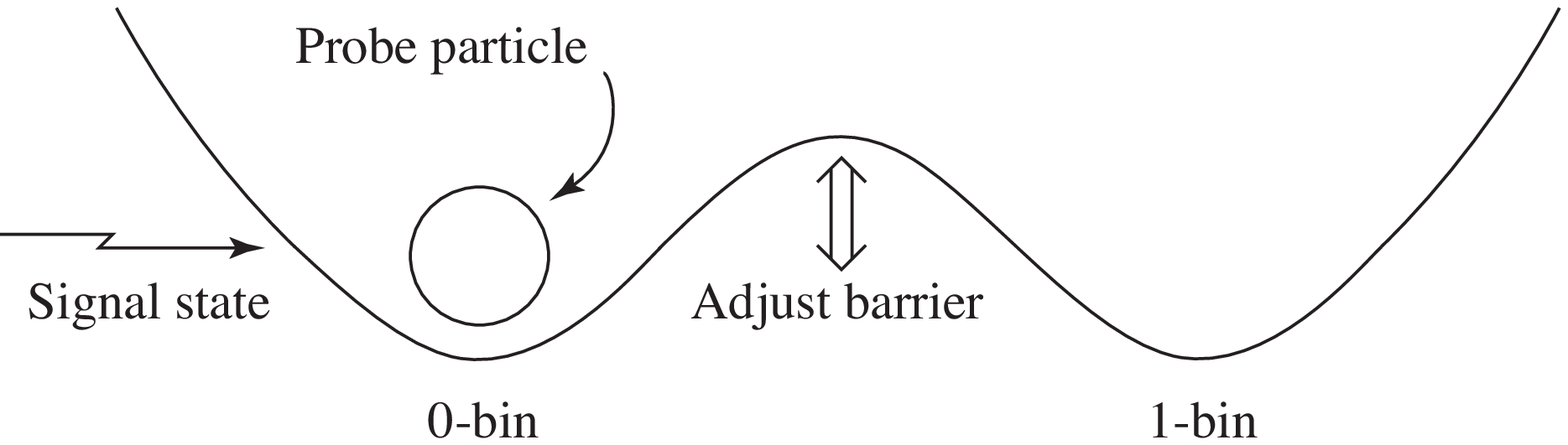}  
   \end{center}
\vspace{50pt}
\caption{Detector as probe particle in a double well.}\label{fig:1}
   \end{figure}

\newpage
\clearpage
\eject
\begin{figure}[t]
\begin{center}
   \includegraphics[width=4.5in]{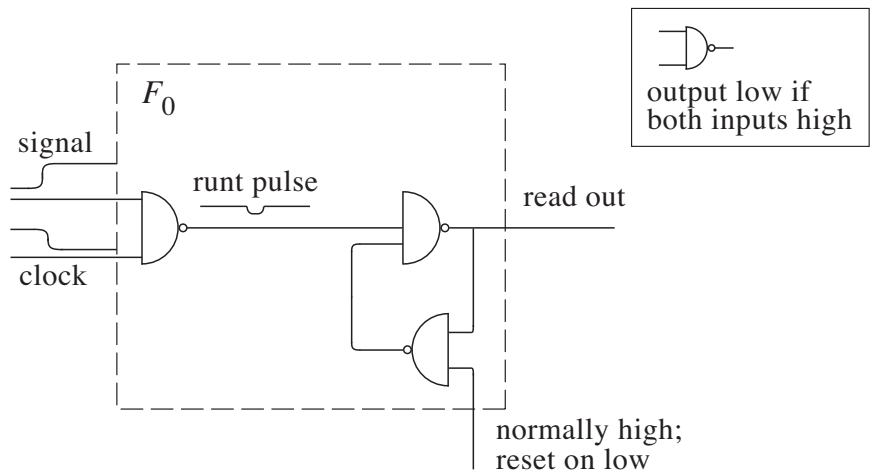} 
   \end{center}
\vspace{50pt}
\caption{Flip-flop exposed to race between signal going high and clock
going low. }\label{fig:2}
   \end{figure}

\newpage
\clearpage
\eject
\begin{figure}[t]
\begin{center}
   \includegraphics[width=4.75in]{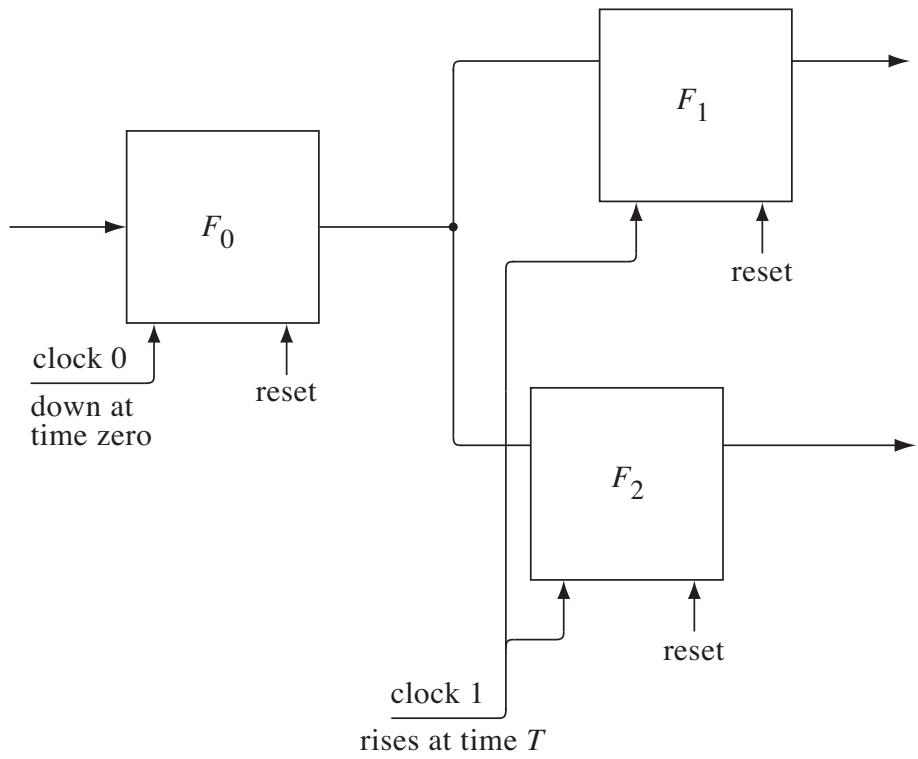}    
   \end{center}
\vspace{50pt}
\caption{Flip-flops $F_1$ and $F_2$ used to read $F_0$ after a delay
$T$. }\label{fig:3}
   \end{figure}

\newpage
\clearpage
\eject
\begin{figure}[t] 
\begin{center}
   \includegraphics[width=5in]{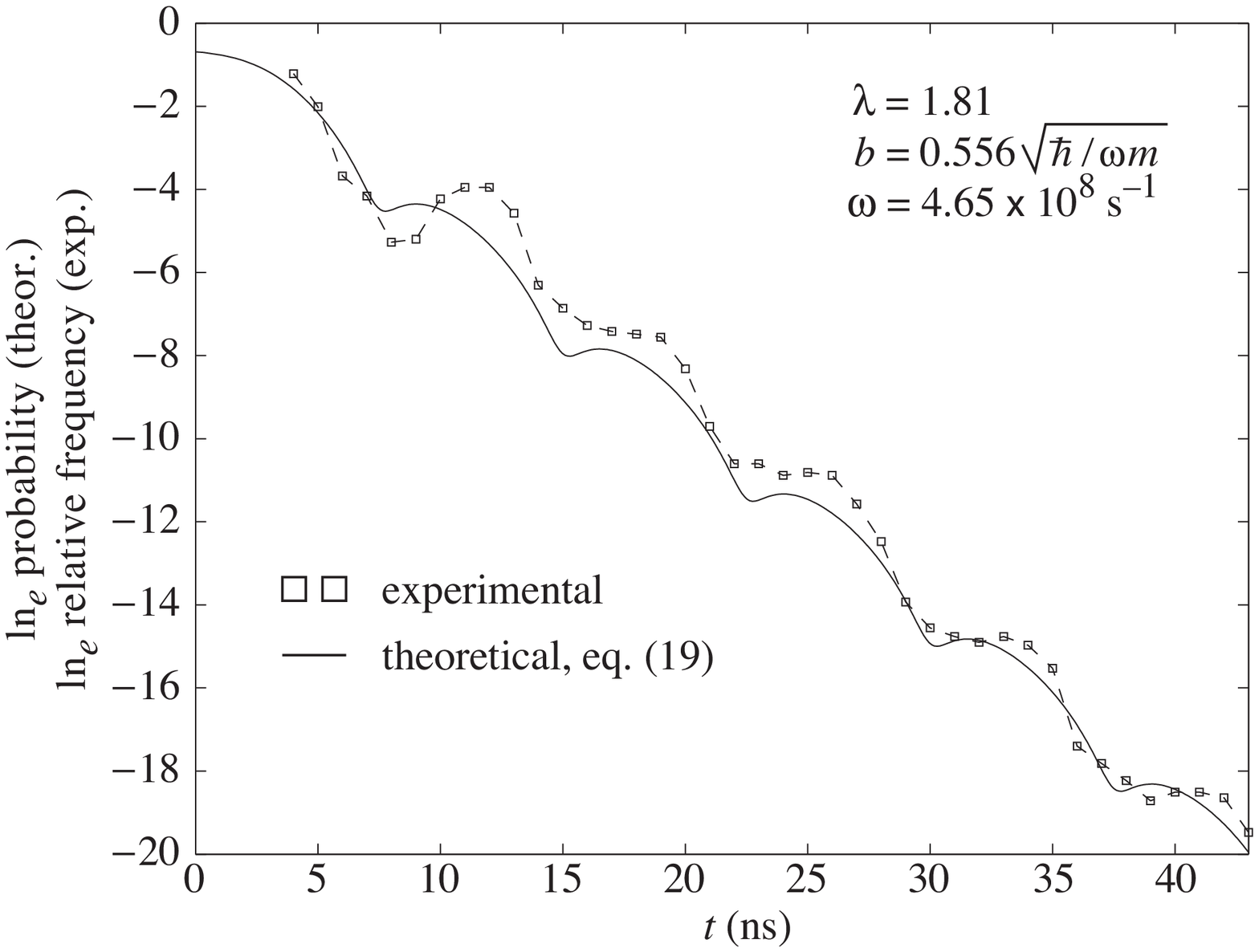}
   \end{center}
\vspace{50pt}
\caption{Probability of disagreement vs.\ settling time.}\label{fig:4}
   \end{figure}


\begin{thebibliography}{15}

\bibitem{dirac} 
P.~A.~M. Dirac, The Principles of Quantum Mechanics, 4th ed.,
Clarendon Press, Oxford, 1958. 

\bibitem{vN} 
J. von Neumann, Mathematical Foundations of Quantum Mechanics
(translated by R.~T. Beyer), Princeton University Press, Princeton,
1955.

\bibitem{rudin} 
W. Rudin, Real and Complex Analysis, 3rd ed., McGraw-Hill, New
York, 1966.

\bibitem{rudin2} 
W. Rudin, Functional Analysis, 2nd ed., McGraw-Hill, New York, 1973.

\bibitem{ams} 
J.~M. Myers and F.~H. Madjid, in:
Quantum Computation and Information, S.~J. Lomonaco, Jr.\
and H.~E. Brandt (Eds.), Contemporary Mathematics Series, Vol.\ 305,
American Mathematical Society, Providence, 2002, pp.\ 221--244.

\bibitem{JOptB}  
J.~M. Myers and F.~H. Madjid, J. Opt.\ B: Quantum Semiclass.\ Opt.\
4 (2002), S109.

\bibitem{aharonov}Y. Aharonov and D.~Z. Albert, Phys.\ Rev.\ D 24 (1981), 359.

\bibitem{tai} 
T. T. Wu, personal communication.

\bibitem{feller}W. Feller, An Introduction to Probability
  Theory and Its Applications, Vol I, 3rd ed., Wiley \&
  Sons, New York, 1968.

\bibitem{sobo} A. Verevkin, G.~N. Gol'tsman, and R. Sobolewski,
 in: OPTO-Canada: SPIE Regional Meeting on Optoelectronic, Photonics, and
Imaging, Technical Digest of SPIE, Vol.\ TD01, SPIE, Bellingham, WA, 2002, pp.\
39--40. 

\bibitem{feynmanIII}R. Feynman, Lectures on Physics, Vol III, Addison-Wesley,
Reading, MA, 1965, chap.\ 7, p.~10. 

\bibitem{gray} 
H.~J. Gray, Digital Computer Engineering,
Prentice Hall, Englewood Cliffs, NJ, 1963, pp.\ 198--201.

\bibitem{chaney}
T.~J. Chaney and C.~E. Molnar, IEEE Trans.\ Computers C-22 (1973),
421.

\bibitem{barton} 
G. Barton, Annals of Physics (NY) 166 (1986), 322.

\end{thebibliography}
\end{document}